\documentclass[twocolumn]{IEEEtran}
\usepackage{url,amsmath,graphicx}

\usepackage{subcaption}
\usepackage{graphicx}
 \usepackage{stfloats}
 \usepackage{eqparbox}
  \usepackage{array}
  \usepackage{fixltx2e}
  \usepackage{stfloats}
   \usepackage{cite}
   \usepackage{url}
\usepackage{fancyhdr}
\usepackage{cite}
\usepackage{graphicx}
\usepackage{psfrag}
\usepackage{url}
\usepackage{stfloats}
\usepackage{amsmath}
\usepackage{array}
\usepackage{fancyhdr}

\usepackage{mathtools, cuted}
\usepackage{lipsum, color}

\usepackage{algorithmicx}
\usepackage{algorithm,algpseudocode}

\makeatletter
\def\BState{\State\hskip-\ALG@thistlm}
\makeatother

\usepackage{amssymb}
\usepackage{color}

\usepackage{tabularx,booktabs}
\begin{document}
%
\title{ Distributed Massive MIMO System with Dynamic Clustering in LEO Satellite Networks
\vspace{-0.2cm}
}
\author{Khaled Humadi\textsuperscript{*},
 Gunes Karabulut Kurt\textsuperscript{*}, and Halim Yanikomeroglu\textsuperscript{**}\\
 \textsuperscript{*}Poly-Grames Research Center, Department of Electrical Engineering, Polytechnique Montreal, Montreal, Canada\\
 \textsuperscript{**}Department of Systems and Computer Engineering, Carleton University, Ottawa, Canada\\
 Emails: khaled.humadi@polymtl.ca, gunes.kurt@polymtl.ca, and  halim@sce.carleton.ca
 \vspace{-0.6cm}
 }
\maketitle

\begin{abstract}
Distributed massive multiple-input
multiple-output (mMIMO) system for low Earth orbit (LEO)  satellite networks is introduced as a promising technique to provide broadband connectivity. 
Nevertheless, several challenges persist in implementing distributed mMIMO systems for LEO satellite networks. These challenges include providing scalable
massive access implementation as the system complexity increases with network size.
Another challenging issue is the asynchronous arrival of signals at the user terminals due to the different propagation delays among distributed antennas in space, which destroys the coherent transmission, and consequently degrades the 
system performance. In this paper, we propose a scalable distributed mMIMO
system for LEO satellite networks based on dynamic user-centric clustering. Aiming to obtain scalable implementation, new algorithms for initial cooperative access, cluster selection, and cluster handover are provided.    In addition, phase shift-aware precoding is implemented to compensate for the propagation delay phase shifts. 
The performance of the proposed user-centric distributed mMIMO is compared with two baseline configurations: the non-cooperative transmission systems, where each user connects to only a single satellite, and the full-cooperative distributed mMIMO systems, where all satellites contribute serving each user.  The numerical results show the potential of the proposed distributed mMIMO system to enhance system spectral efficiency when compared to non-cooperative transmission systems. Additionally, it demonstrates the ability to minimize the serving cluster size for each user, thereby reducing the overall system complexity in comparison to the full-cooperative distributed mMIMO systems.
\\
\\
\end{abstract}
\vspace{-0.6cm}
\vspace{-0.6cm}

%
\IEEEpeerreviewmaketitle

\section{Introduction}
Multiple-input multiple-output (MIMO) is a technology implemented to enhance the performance in wireless communication systems \cite{Wangcellular}. However, co-located MIMO in satellites communication fails to provide the same performance as that in terrestrial networks \cite{ArapoglouMIMO}. This is because satellite channels generally differ from those of the terrestrial network. In particular, strong line-of-sight (LoS) channel is one of the main characteristics of satellite channels due to the lack of scatters in the space. This satellite channel characteristic results in high spatial correlation when co-located MIMO systems are used, making it difficult for the receiver to estimate the channel coefficient for each antenna and hence affecting the validity of the channel state information which is necessary for MIMO system operation. 
As such, satellite networks, have the drawback in that they are unable to effectively exploit the performance enhancement provided by MIMO technology compared to terrestrial networks \cite{HeoMIMO}. To overcome this issue, instead of using co-located MIMO, a distributed MIMO system is proposed which is able to provide uncorrelated channels among spatially separated antennas \cite{AbdelsadekDistributed}. 

{Distributed MIMO in space requires the cooperation of satellites to enable beamforming or joint transmission and reception. In \cite{RichterDownlink},  a joint transmission scheme is investigated using two LEO satellites. It is shown that the relative positioning of the cooperating satellites and the ground user substantially impacts system performance due to the predominance of LOS channel.  The authors in \cite{SyangD}
introduced orthogonal time frequency space (OTFS) for distributed multi-satellite cooperative transmission scheme to effectively enhance downlink performance. 
The problem of overlap in satellite coverage areas is studied in \cite{ArtiData} where a user receives multiple signals in a multi-satellite scenario. In \cite{AbdelsadekDistributed} a distributed massive MIMO (mMIMO) system is proposed for LEO satellite networks where multiple satellite access points cooperate to serve a set of ground users. 
A theoretical framework to investigate the spectral efficiency performance of distributed mMIMO  in space was developed in \cite{abdelsadek2023broadband}. The results in \cite{AbdelsadekDistributed} and \cite{abdelsadek2023broadband} show that in addition to improving the average system spectral efficiency, distributed mMIMO can reduce the handover rate since the user receives its data from a cluster of serving satellites instead of a single satellite connection. In \cite{XuEnhancement}, a distributed beamforming technique was proposed to support direct communications from LEO satellites to smartphones. To enhance the user quality of service (QoS), a coordinated multi-satellite joint transmission is investigated in \cite{XLiB}.    The study demonstrates that this system not only achieves higher coverage probability than traditional single-satellite systems but also optimizes user's ergodic rates.}

To effectively exploit this technology in satellite networks, several issues need to be addressed.  For example, due to the increase in the density of LEO satellite constellation and the high dynamics of satellites, utilizing static-clustering cooperation significantly increases the system complexity with increasing the network size. Furthermore, to achieve such cooperation in space, satellites need to be synchronized in time, frequency, and phase. Although these issues are well addressed in terrestrial networks, as well as GEO satellite networks, the current growth of the LEO satellite networks and their operational implementation necessitate extensive research that accurately captures the distinctive attributes and benchmarks of LEOs.

In this paper, we propose a scalable distributed mMIMO system designed for LEO satellite networks, utilizing dynamic user-centric clustering. The proposed system is called a user-centric distributed mMIMO (UC-DMIMO) system.  With a focus on achieving scalable implementation, we provide new algorithms for initial cooperative access, pilot assignment, and cluster selection. 
To further enhance system efficiency  in terms of complexity and handover rate, each user is allocated a reference satellite access point (RSAP) responsible for exchanging the user's information with other satellites in its designated serving cluster (SC). 
Furthermore, phase-aware pre-coding is used to synchronize the phases of signals received from different spatially distributed antennas.  Finally, the system performance in terms of spectral efficiency and the SC size of the proposed UC-DMIMO system is compared with that of the baseline full-cooperative DMIMO (FC-DMIMO) systems, where all satellites cooperatively serve all users simultaneously, as well as non-cooperative transmission (NCT) systems, where each user is connected to only a single satellite. 



The rest of this paper is organized as follows. In Section II, we introduce the architecture of the UC-DMIMO LEO satellite networks, where several aspects of the architecture are addressed, i.e., channel model, pilot assignment, and Dynamic UC satellite clustering. 
In Section III, we discuss
the downlink transmission of the proposed architecture. In Section IV,  the
simulation results are presented to investigate the performance of the
proposed UC-DMIMO.
Finally, Section V concludes the work in this paper.





\vspace{-0.2cm}
\section{System Model}
\vspace{-0.2cm}
\subsection{Network Model}
\begin{figure}
\centering
\begin{subfigure}
{0.4\textwidth}
\includegraphics[width=\textwidth]{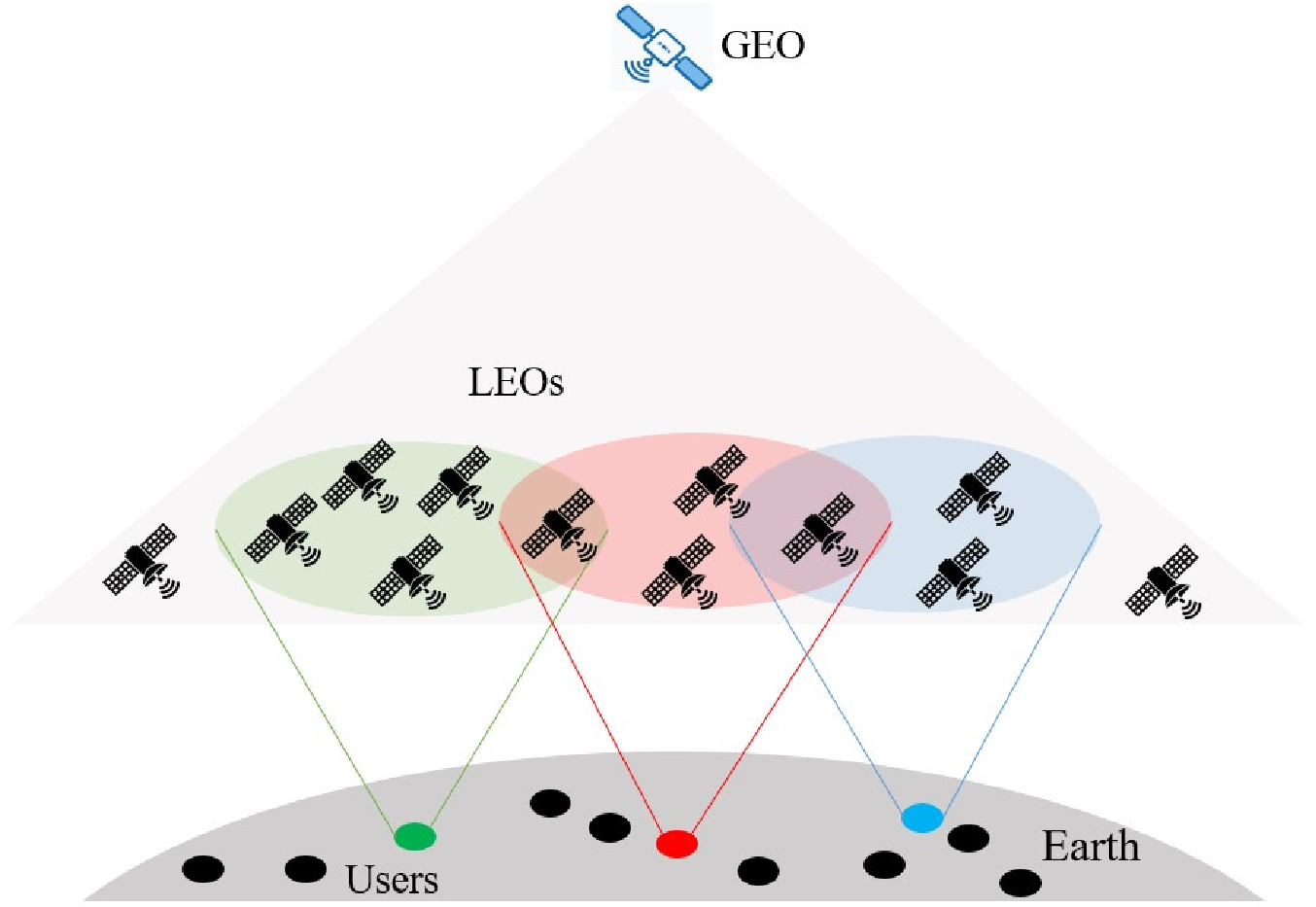}
    \caption{}
    \label{SMa}
\end{subfigure}
\vspace{0.3cm}
\begin{subfigure}{0.3\textwidth}
\vspace{0.2cm}
\includegraphics[width=\textwidth]{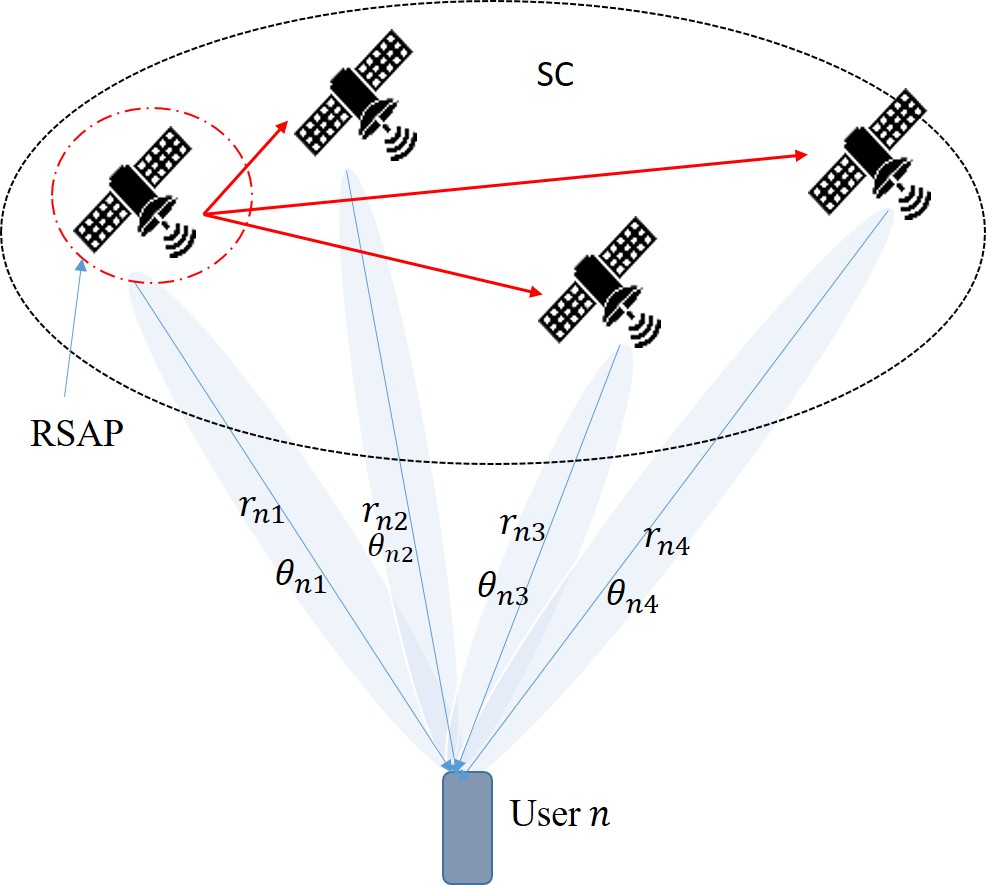}
    \caption{}
    \label{SMb}
\end{subfigure}   
\vspace{-0.3cm}
\caption{UC-DMIMO system in space. a) general system with a GEO satellite working as a CPU and a set of orbiting LEO satellites working as access points that jointly serve ground users; and b)  the $n$-th SC of LEO satellites selected based on a user-centric clustering approach and cooperatively work as a distributed antenna system to serve the $n$-th user.  \vspace{-0.3cm}}
\label{SM}
\vspace{-0.3cm}
\end{figure}

In this paper, we consider a  downlink satellite network, where $M$ spatially distributed and orbiting LEO satellites are serving $N$ spatially distributed ground users. Each user is cooperatively served by a subset $\mathcal{A}\subseteq M$ LEO Satellites in order to provide reliable, continuous, and efficient data transmission. Each LEO satellite is assumed to be equipped with $L$ antennas, while terrestrial users are single-antenna terminals. This satellite cooperation performs as a UC-DMIMO system in space, in which each user carefully selects its SC of satellites. 
The network architecture is depicted in Fig. \ref{SMa}. Backhaul links connect the LEO satellites to a GEO satellite which works as a 
central processing unit (CPU)  conveying necessary control signals between them. Although backhauling system constraints such as propagation delay and link capacity are important issues, we focus in this paper on the access lines, i.e., links between LEO satellites and ground users.
Furthermore, we consider the time-division duplex protocol, where each coherence block  consists of $\tau_c$ time instants (channel uses) with $\tau_p<\tau_c$ is assigned for the uplink training phase 
 and $\tau_c-\tau_p$ for the downlink data transmission phase.

 \vspace{-0.24cm}
\subsection{Channel Model}
The
channels between the satellites and ground users consist of Los and
non-LoS (NLoS) links, where the probability of the link being LoS increases with the satellite elevation angle, reaching its maximum at an elevation angle of $90^o$  \cite{study3GPP}. Therefore, the channel between the   $n$-th user and the $m$-th satellite, denoted by $\textbf{h}_{mm}\in \mathbb{C} ^{L\times 1}$  is modelled as Rician which is given as
\begin{equation}
\textbf{h}_{nm}=\beta_{nm}\textbf{g}_{nm},
    \label{chM}
\end{equation}
where $\beta_{nm}$ represents the large-schale fading and $\textbf{g}_{nm}\in\mathbb{C} ^{L\times 1}$ is the small-scale fading. For $\textbf{g}_{nm}$, we have 
\begin{eqnarray}
\textbf{g}_{nm}=\sqrt{\frac{\kappa}{1+\kappa}} \textbf{g}^\prime_{nm} +\sqrt{\frac{1}{1+\kappa}} \textbf{g}^{\prime\prime}_{nm},  
\end{eqnarray}
where $\kappa$ refers to the Rician K-factor and 
$\textbf{g}^\prime_{nm}$ and $\textbf{g}^{\prime\prime}_{nm}$ corresponds to the LOS and NLOS components. Specifically, $\textbf{g}^{\prime\prime}_{nm}$ characterizes the random component, accounting for the scattered multipath that follows a Rayleigh distribution its elements are independent and identically distributed random variables with a zero-mean and unit variance. Besides, according to \cite{JZHANGPerformance}, the $l$-th element of  $\textbf{g}^\prime_{nm}$ is given as 
\begin{eqnarray}  
{{g}^\prime}_{nm}^l=e^{-j(l-1)2\pi d/\lambda \sin{\vartheta_{nm}}},
\end{eqnarray}
where $d$ is the antenna
spacing, $\lambda$ represents the signal wavelength, and $\vartheta_{nm}\sim \mathcal{U}\big[0, 2\pi\big]$ denotes the angle of arrive.
In (\ref{chM}), the large-scale fading of the channel between the  $n$-th user and the $m$-th LEO satellite, $\beta_{nm}$  is defined as\cite{AbdelsadekBroadBand}
\begin{equation}
    \beta_{nm}=\frac{1}{{\mathcal{L}^{prop}_{nm}}\mathcal{L}^{sh}_{nm}\mathcal{L}^{ant}_{nm}\mathcal{L}^{other}_{nm}},
    \label{pathloss}
\end{equation}
where ${\mathcal{L}^{prop}_{ml}}$ is the free-space propagation path-loss which is a function of the operating frequency $f$ and the distance $r_{nm}$ and can be obtained in dB as
\begin{equation}
   {\mathcal{L}^{prop}_{nm}} [\textrm{dB}]=32.45+20\textrm{log}_{10}f+20\textrm{log}_{10}r_{nm}.
\end{equation}
The parameters  $\mathcal{L}^{sh}_{nm}$ in (\ref{pathloss}) represents the 
losses due to the shadowing effect, modeled as a log-normal random variable, i.e., $\mathcal{L}^{sh}_{nm}\sim\mathcal{N}(0,\sigma_{sh}^2)$, where $\sigma_{sh}^2$ is the shadowing variance. $\mathcal{L}^{ant}_{ml}$ denotes the gain losses due to beam misalignment . Based on the 3GPP report in \cite{study3GPP}, this losses is modeled as
\begin{equation}
    \mathcal{L}^{ant}_{nm}=\frac{1}{4}\bigg|\frac{2\pi \eta \sin{w_{nm}}}{J_1(2\pi \eta \sin{w_{nm}})}\bigg|^2,
\end{equation}
where $J_1(\cdot)$ is the  first kind and first
order Bessel function, $w_{nm}$ is boresight angle of the
$m$-th satellite’s antenna beam with respect to the $n$-th user, and $\eta$ represents the aperture radius of the antenna in wavelengths. Finally, $\mathcal{L}^{other}_{nm}$  in (\ref{pathloss}) accounts for the other large-scale losses such 
as atmospheric gasses attenuation and ionospheric scintillation \cite{study3GPP}. 
\vspace{-0.3cm}
\subsection{Asynchronous Reception}
In distributed antenna systems, the main causes of the asynchronous reception are 
the variances in the propagation delays and oscillator hardware imperfection. More specifically, these factors result in a multiplicative random phase in the transmitted signal. In this work, we focus on the phase shift resulting from the variant propagation delays.  
  Since the distances between spatially distributed LEO satellites in a given user's SC are very long and random, the propagation time of the transmitted signals from all serving LEO satellites to a user is different which results in asynchronous signal
arrival. Therefore, with respect to the phase of the signal received from the RSAP, each other received signal at the user experiences a phase shift related  to the relative
position of the satellite and the user \cite{YanAsynchronous,JImpacts,JAsynchronous}. As such, the phase shift in the channel between the $m$-th satellite and the $n$-th user is expressed as 
\begin{equation}
\theta_{nm}=\exp{\big(-2\pi\frac{\Delta t_{nm}}{T_s}\big)},
\label{phaseshift}
\end{equation}
where $T_s$ denotes the symbol duration, $\Delta t_{nm}= \frac{\Delta r_{nm}}{c}$ is the timing offset of the signal transmitted from the $m$-th satellite to the $n$-th user with  $\Delta r_{nm}$ and $c$ are the distance difference and light speed, respectively. 
To this end, taking into consideration the phase shift effect, the channel between the $n$-th user and $m$-th satellite is written as
\begin{equation}
   \Tilde{\bold{h}}_{nm}=\theta_{nm}\textbf{h}_{nm}=\theta_{nm} \beta_{nm}\textbf{g}_{nm}.
\end{equation}

The phase shift $\theta_{mn}$ is mainly determined by the positions of satellite $m$ and user $n$ at the initial access and then updated dynamically based on the predicted satellite mobility. 
\vspace{-0.3cm}
\section{User-centric clustering and downlink transmission}
\subsection{Cluster Formation and Reference Satellite}
In the proposed UC-DMIMO system, each user is connected to a set of satellites in its vicinity. The user vicinity is determined by a given minimum elevation angle. 
To provide flexible and scalable configuration, each user is assigned a reference satellite access point (RSAP) based on a given criteria. In addition to serving the user, the RSAP is responsible for assigning resources and changing the user's information with other satellites in its SC including the user's location and assigned pilot sequence. This configuration will reduce overhead and hence the overall system complexity. In this paper, we introduced two criteria to determine the user's RSAP: a) best channel-based RSAP and b) maximum service time-based RSAP. Since the orbital dynamics of satellites can be predicted, the serving time is assumed available. Let $\zeta_{nm}$ and $\mathcal{S}_n$, respectively, denote the serving time of user $n$ from satellite $m$ and the set of satellites with a predetermined minimum elevation angle with respect to the $n$-th user. 
Consider a new parameter, $\Lambda_{nm}$ defined for user $n$ and satellite $m$ such that $\Lambda_{nm}=\beta_{nm}$ if the RSAP is selected based on the best average channel gain and  $\Lambda_{nm}=\zeta_{nm}$ if the RSAP is selected based on the maximum service time. In the following, we provide two algorithms, namely Algorithm 1 and Algorithm 2, for initial SC formation as well as SC update and handover. \vspace{-0.5cm}
\begin{algorithm}[t]
\caption{Initial user access and SC selection }
\label{alg:description}

\textbf{Input:} Candidate SCs ${\cal S}_n$,  average channel gain $\beta_{nm}$, and serving time $\zeta_{nm}$ between user $n$ and satellite $m\in {\cal S}_n$  for all $n\in\{1,2,..., N\}$. 
\\
\textbf{Output:} Each user's SC $\mathcal{A}_n$ and $RSAP_n$ for all $n\in\{1,2,...,N\}$.
\begin{algorithmic}[1]
    \For{$n = 1$ to $N$}
        \State An RSAP is selected for user $n$ from ${\cal S}_n$ such that \textcolor{white}{ 1111} $RSAP_n=\{\textrm{satellite}\; m\!\!: m=\arg \max_{m\in\mathcal{S}_n}{\Lambda_{nm}}$\}
        \State User $n$ sends a serving request to the $RSAP_n$ 
        \If{$RSAP_n$ deny the user's serving request} 
            \State set ${\cal S}_n \gets {\cal S}_n \setminus \{RSAP_n\}$.
            \State go to Step 2
        \Else
            \State $\mathcal{A}_n \gets \{RSAP_n\}$ 
            \State pilot sequence $\tau_n$ is assigned to user $n$.
            \State User $n$ locks its phase to the $RSAP_n$.
        \EndIf
        \State User $n$ sends its ${\cal S}_n$ and location to $RSAP_n$.
        \State The $RSAP_n$ informs the user's serving request  to all \textcolor{white}{ 111} satellites in $
        {\cal S}_n$. 
        \If{ the pilot $\tau_n$ in satellite $i\in {\cal S}_n$ is available}, 
        \State{Satellite $i$ will join the user's SC, i.e., \textcolor{white}{ 1111111}$\mathcal{A}_n \gets \mathcal{A}_n\bigcup\{\textrm{satellite}\; i\}$}
        \EndIf
    \EndFor
\end{algorithmic}
\end{algorithm}
\begin{algorithm}[t]
\caption{SC update and handover }
\label{alg:description}

\textbf{Input:} ${\cal A}_n$ and ${\cal S}_n$  for each user $n\in\{1,2,..., N\}$.
\\
        \textbf{Output:} Updated 
        $\mathcal{A}_n$ for all $n\in\{1,2,...,N\}$.
\begin{algorithmic}[1]
    \For{$n = 1$ to $N$}
    \If{$RSAP_n\in {\cal S}_n$}
    \State User $n$  updates its location and ${\cal S}_n$  to the $RSAP_n$.
    \If{New satellite $i$ join $S_n$}
    \State The $RSAP_n$ informs the user's serving request \textcolor{white}{ spacespace} to satellites $i$. 
        \If{ the pilot $\tau_n$ in satellite $i\in {\cal S}_n$ is available \textcolor{white}{ space1-space2}},  
        \State{$\mathcal{A}_n \gets \mathcal{A}_n\bigcup\{\textrm{satellite}\; i\}$}
        \EndIf
        \EndIf
        \If{Satellite $j\in\mathcal{A}_n\;\&\; j\notin\mathcal{S}_n$}
        \State{$\mathcal{A}_n \gets \mathcal{A}_n\backslash \{\textrm{satellite}\; j\}$}
        \EndIf
        \Else
        \State Cluster handover occurs.
        \EndIf
        \EndFor
\end{algorithmic}
\end{algorithm}
\subsection{Channel Estimation}
Each satellite is required to estimate uplink channels from all users utilizing the uplink pilot sequences. We consider $\tau_p$ mutually orthogonal time-multiplexed pilot
sequences are used. Since each satellite is supported with $L$ antennas, we assume the pilots assigned for these $L$ antennas are orthogonal, i.e., there is no pilot reuse among the co-located antennas. However, pilot sequences can be reused for other distributed antennas in the network. 
Let $\bold{\Xi}_n\in\mathbb{C}^{\tau_p}$ denotes the $L$ pilot sequences assigned to
the $n$-th user where $||\bold{\Xi}_n||^2=L\tau_p$ and let
 $\mathcal{C}_n$ represents the subset of users that use the pilot sequences $\bold{\Xi}_n$ as the $n$-th user. 
 Then, the received pilot signal $\bold{Y}^P_{m}\in\mathbb{C}^{L\times \tau_p}$ at the  $m$-th satellite is given by
\begin{equation}
\bold{Y}^P_{m}=\sum_{n=1}^N \sqrt{a_{nm} p_n}\Tilde{\bold{h}}_{nm}\bold{\Xi}_n^T+\bold{W}_m,  
\end{equation}
where $p_n$ is the pilot transmit power of user $n$, $\bold{W}_m\in \mathbb{C}^{L\times \tau_p}$ is the additive thermal noise matrix which has i.i.d elements $w_m^{ij}\sim \mathcal{N}\big(0,\sigma^2\big)$, and $a_n$ is an indicator parameter to determine if the $m$-th satellite belongs to the $n$-th user's SC, $\mathcal{A}_n$, which can be defined as
\begin{eqnarray}
     a_{nm}=\left\{
     \begin{array}{ll}
     1, \quad\quad\quad m\in \mathcal{A}_n \\
     0, \quad\quad\quad   \textrm{otherwise} 
\end{array}
\right.
\label{GModel}
\end{eqnarray}
Therefore 
the required information to estimate the channel of the $n$-th user is given as 
\begin{equation}
\bold{y}^{\textrm{p}}_{nm}=\frac{\bold{Y}^P_{m}\bold{\Xi}^*_n}{\sqrt{\tau_p}}=\sqrt{\tau_p}\sum_{j\in\mathcal{C}_n} \sqrt{a_{jm} p_j}\Tilde{\bold{h}}_{jm}+\bold{w}_m,  
\end{equation}
where $\bold{w}_m=\bold{W}_m\bold{\Xi}^*_n/\sqrt{\tau_p}\sim \mathcal{N}\big(0,\sigma^2\bold{I}_N\big)$ with $\bold{I}_N$ represents the identity matrix. 
Then, the estimate of the uplink channel from the $n$-th user to the $m$-th satellite, denoted as $\hat{\bold{h}}_{nm}$, is obtained using the linear minimum mean square estimator (LMMSE) \cite{Bjornsonscalable}. As such, $\hat{\bold{h}}_{nm}$ is expressed as
\begin{eqnarray}
 \hat{\bold{h}}_{nm}&=&\frac{\mathbb {E}\Big[\bold{h}_{nm}\bold{y}_{nm}^H\Big]}{\mathbb {E}\Big[\bold{y}^{\textrm{p}}_{nm}(\bold{y}^{\textrm{p}}_{nm})^H\Big]} \bold{y}_{nm}^{\textrm{p}}\nonumber\\
 &&\!\!\!\!\!\!\!\!\!\!=\frac{\sqrt{p_n a_{nm} }\theta^*\bold{R}_{nm}}{\sum_{j\in \mathcal{C}_n}{p_{j}a_{jm}}\bold{R}_{jm}+\sigma^2\bold{I}_L} \bold{y}_{nm}^{\textrm{p}},
 \label{estmator}
\end{eqnarray}
where $\bold{R}_{nm}=\beta_{nm}\textbf{g}^\prime_{nm}(\textbf{g}^\prime_{nm})^H+\beta_{nm}\bold{I}_L$ is the spatial correlation matrix of the channel between the $n$-th user and the $m$-the satellite. It is clear from (\ref{estmator}) that the square matrix $\frac{\sqrt{p_n }\bold{R}_{nm}^H}{\sum_{j\in \mathcal{C}_n}{p_{j}a_{jm}}\bold{R}_{jm}+\sigma^2\bold{I}_L}\in \mathbb{C}^{L\times L}$ is a function of the channel statistics which can be predicted. This means that channel estimation can be performed locally in each satellite. 


\vspace{-0.3cm}
\subsection{Downlink Transmission}
We assume that a coherent
joint transmission is applied in the downlink, which means
that all satellites in the user's SC send the same data symbols. As such the transmitted signal vector from the $m$-th  satellite can be expressed as
\begin{eqnarray}
\textbf{x}_m=\sum_{n=1}^{N} \sqrt{\varkappa_{nm}a_{mn}}\textbf{v}_{nm}s_n,   
\end{eqnarray}
where $s_n\sim \mathcal{N}(0,1)$ is the transmitted symbol to user $n$, $\varkappa_{nm}$ is the power allocation coefficient from the $m$-th satellite to the $n$-th
 user, and $\textbf{v}_{nm}$ is the precoding vector utilized by the $m$-th satellite to precode the date transmitted to user $n$. Then, the received signal at the $n$-th user is given as
\begin{eqnarray}
    y_n&=&\sum_{m=1}^M\textbf{h}_{nm}^H\textbf{x}_{nm}s_n=\sum_{m=1}^M\sqrt{\varkappa_{nm}a_{nm}}\textbf{h}_{nm}^H\textbf{v}_{nm}s_n\nonumber\\
    &&+ \sum_{m=1}^M\sum_{\substack{i=1\\ i\neq n}}^N\sqrt{\varkappa_{im}a_{im}}\textbf{h}_{im}^H\textbf{v}_{im}s_i+w_n,
\end{eqnarray}
where $w_n$ represents the receiver thermal noise at user $n$. Utilizing maximum ratio processing allows for leveraging the distributed channel estimation at the satellites. This stands out as a key advantage of the distributed antenna system, with the potential to lower both computational complexity and the necessary backhaul signaling between the LEO satellites and the CPU \cite{CellfreeLarsson}. As such, the the precoding vector of the
$m$-th satellite  to the $n$-th user is 
\begin{eqnarray}
    \textbf{v}_{nm}=\theta_{nm} \hat{\bold{h}}_{nm}.
\end{eqnarray}

Therefore, the signal-to-interference-plus-noise ratio (SINR) at the $n$-th user can be  expressed as
\begin{eqnarray}
    \textrm{SINR}_n=\frac{\Big| \mathbb {E}\Big\{\sum_{m=1}^M\sqrt{\varkappa_{nm} a_{nm}}\textbf{h}_{nm}^H\textbf{v}_{nm}\Big\}\Big|^2}{ \sum_{m=1}^M {\sum_{\substack{i=1\\ i\neq n}}^N \varkappa_{im}a_{im}}\mathbb {E}\Big\{\Big|\textbf{h}_{im}^H\textbf{v}_{im}\Big|^2\Big\}+w_n}.
\end{eqnarray}
Thus, the achievable
spectral efficiency for the $n$-th user is 
\begin{eqnarray}
    \textrm{SE}_n&=&\big(1-\frac{\tau_p}{\tau_c}\big) \log_2\big(1+\textrm{SINR}_n\big)=\big(1-\frac{\tau_p}{\tau_c}\big)\nonumber\\
    &&\!\!\!\!\!\!\!\!\!\!\!\!\!\!\!\!\!\!\!\!\!\!\!\!\!\!\!\!\!\!\!\!\!\!\!\times \log_2\big(1+\frac{\Big| \mathbb {E}\Big\{\sum_{m=1}^M\sqrt{\varkappa_{nm} a_{nm}}\theta_{nm}\textbf{h}_{nm}^H\hat{\bold{h}}_{nm}\Big\}\Big|^2}{E_n+F_n+w_n}\big),
\end{eqnarray}
where $E_n$, and $R_n$, respectively, are the beamforming uncertainty gain, and multi-user interference and are given by
\begin{eqnarray}
    E_n=\textrm{var}\Big\{\sum_{m=1}^M\sqrt{\varkappa_{nm} a_{nm}}\theta_{nm}\textbf{h}_{nm}^H\hat{\bold{h}}_{nm}\Big\},
\end{eqnarray}
\begin{eqnarray}
    F_n= \sum_{m=1}^M {\sum_{\substack{i=1\\ i\neq n}}^N \varkappa_{im}a_{im}}\theta_{nm}\mathbb {E}\Big\{\Big|\textbf{h}_{im}^H\hat{\bold{h}}_{im}\Big|^2\Big\}.
\end{eqnarray}
In a specific scenario when users are dispersed within a confined geographical region, it becomes feasible to regard all users as being served by the same SC, provided there are a sufficient number of orthogonal pilot sequences. 




\vspace{-0.4cm}
\section{Simulation Results}

In this section, simulation results are provided in order to demonstrate the performance of the proposed UC-DMIMO system for LEO satellite networks.  Unless otherwise specified, the system parameters considered in the simulation results are set as follows. The operating frequency is 2GHz,  antenna factor $\eta=10$ wavelengths, shadowing variance $\sigma^2_{sh}=5$ dB, noise power spectral density $\sigma=-174$ dBm/Hz, Rician K-factor, $\kappa=10$, satellite antenna gain and maximum transmit power are 30 dBi and 15 dBW, respectively, user antenna gain 0 dB, pilot power, $p_n=p=0$ dBW, length of coherent time and pilot sequences are $\tau_c=200$, and $\tau_p=30$, respectively, minimum elevation angle is 0, number of users $N=10$, and number of orbiting LEO satellites $M=100$ and $400$.

\begin{figure}
\centering
\begin{subfigure}{0.4\textwidth}    \includegraphics[width=\textwidth]{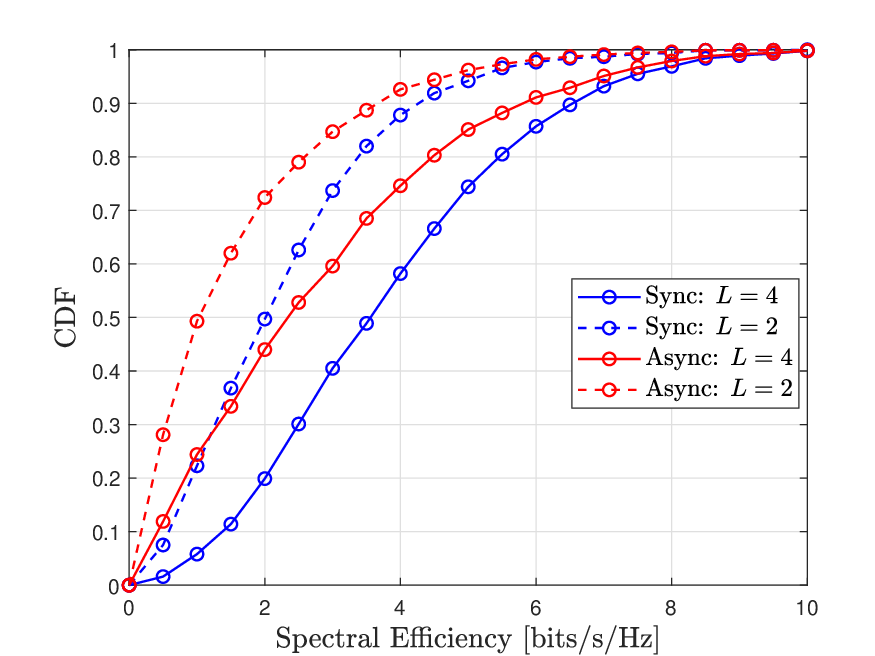}
    \caption{}
    \label{Fig2a}
\end{subfigure}
\hfill
\begin{subfigure}{0.4\textwidth}
\includegraphics[width=\textwidth]{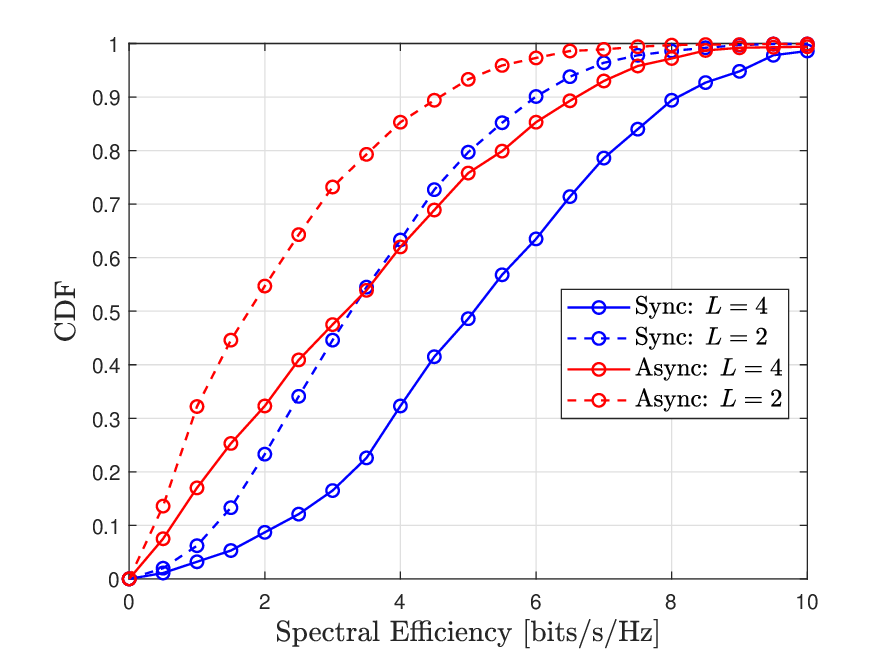}
    \caption{}
    \label{Fig2b}
\end{subfigure} 
\caption{Downlink per-user spectral efficiency 
for a) $M=100$ and b) $M=400$, when the RSAP is selected based on the maximum channel gain. Note that in this paper, we mainly focus on investigating phase synchronization. 
\vspace{-0.6cm}
}
\label{Fig2}
\end{figure}

In Figs. \ref{Fig2a} and \ref{Fig2b}, we plot the cumulative distribution function (CDF) of the per-user downlink spectral efficiency of the proposed UC-DMIM system for $M=100$ and $M=400$, respectively. These figures illustrate the system performance when the RSAP is selected based on the best channel criteria.  As depicted in the figure, since the propagation delay phase can be perfectly estimated 
 and incorporated into the precoding process, the system performance is significantly enhanced when this phase shift is compensated. For example, as shown in Fig. \ref{Fig2b}, the performance of the synchronous transmission with $L=2$ is almost the same as that of asynchronous transmission with $L=4$. This indicates that phase compensation can either enhance the system performance or reduce the number of distributed or co-located antenna requirements. Furthermore, it is clear from Figs. \ref{Fig2a} and \ref{Fig2b} that the system performance in terms of spectral efficiency is enhanced with increasing the number of orbiting satellites since this increases the probability of more satellites joining the user's SC.  

\begin{figure}[t]
    \centering
 \includegraphics[scale=0.5]{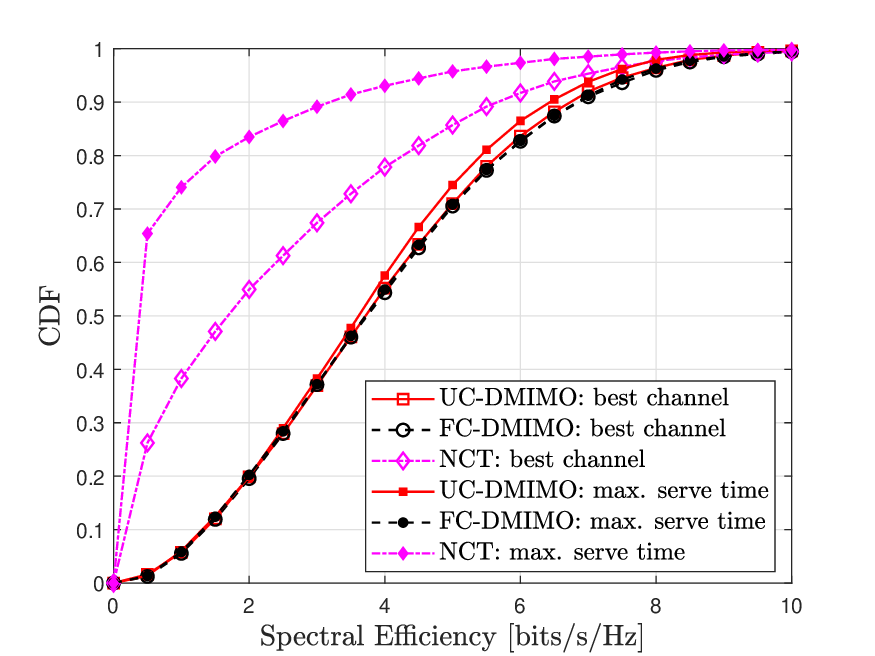}
    \caption{Comparison of the downlink per-user spectral efficiency for the proposed UC-DMIMO system with the FC-DMIMO and the NCT when the RSAP is selected based on the best channel and based on the maximum service time.
    }
    \label{Fig4}
\vspace{-0.6cm}
\end{figure}

In Fig. \ref{Fig4}, we compare the spectral efficiency performance of the proposed UC-DMIMO system with that of the FC-DMIMO \cite{AbdelsadekBroadBand} and the NCT system \cite{Wegraph}. This comparison is conducted under both RSAP selection methods, 
 and for $M=100$ and $L=4$. It is clear that the proposed UC-DMIMO system provides a performance comparable to the FC-DMIMO system while outperforming that of the NCT systems. 
 Furthermore, for the UC-DMIMO  and FC-DMIMO systems, both selection methods provide approximately the same performance in terms of per-user spectral efficiency. 
This is due to the system's design of distributed MIMO, where the user is connected to a cluster of satellites. One of these satellites is designated as the user's RSAP 
and it cooperatively serves the user alongside the other satellites in its SC. This cooperative approach helps mitigate the effects of interference under both selection methods. On the other side, in the NCT systems, each user is connected to a single satellite which is considered the RSAP as well. If this satellite is selected based on the maximum serve time, there may be other satellites closer to the user which results in a high level of interference. As such, selecting the serving satellite in the NCT architecture based on the maximum serve time results in more degraded performance   
as illustrated in Fig. \ref{Fig4}.


\begin{figure}[t]
    \centering
    \includegraphics[scale=0.5]{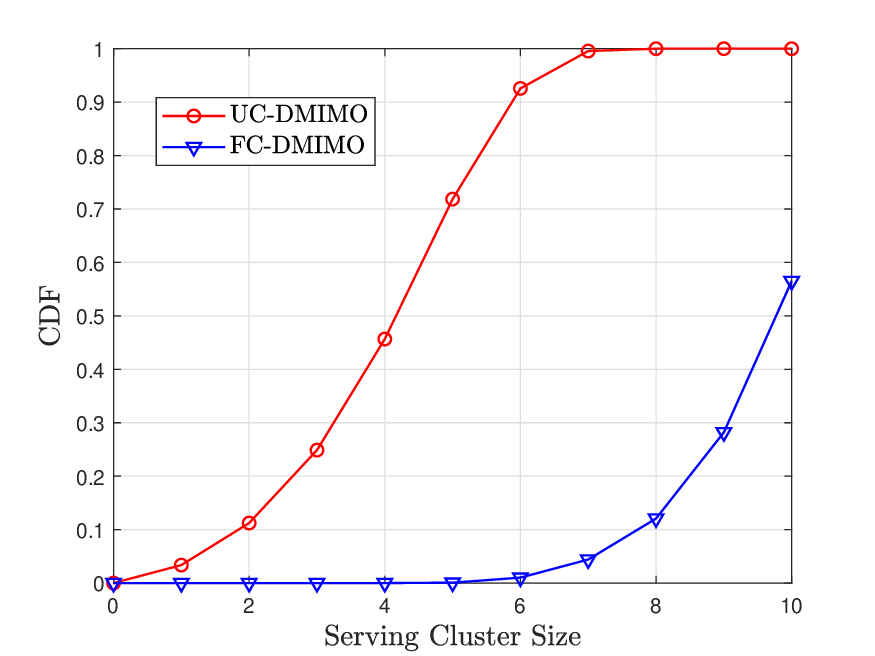}
    \caption{User's SC size for the proposed US-DMIMO compared with that of FC-DMIMO when the RSAP is selected based on the best channel. } 
    \label{Fig7}
   \vspace{-0.5cm}
\end{figure}

The results depicted in Fig. \ref{Fig7} show the distribution of the user's SC size for the proposed UC-DMIMO and compare it with that of the FC-DMIMO. While Fig. \ref{Fig4} demonstrates comparable spectral efficiency performance between UC-DMIMO and FC-DMIMO, Fig. \ref{Fig7} reveals a noteworthy distinction in the user's SC size, i.e., the number of serving satellites. 
Specifically, the user's SC size in UC-DMIMO is significantly smaller than that in FC-DMIMO. This observation indicates that UC-DMIMO offers a reduction in backhaul overhead, leading to a decrease in system complexity and facilitating scalable implementation in cooperative satellite networks.

\begin{figure}[t]
    \centering
    \includegraphics[scale=0.5]{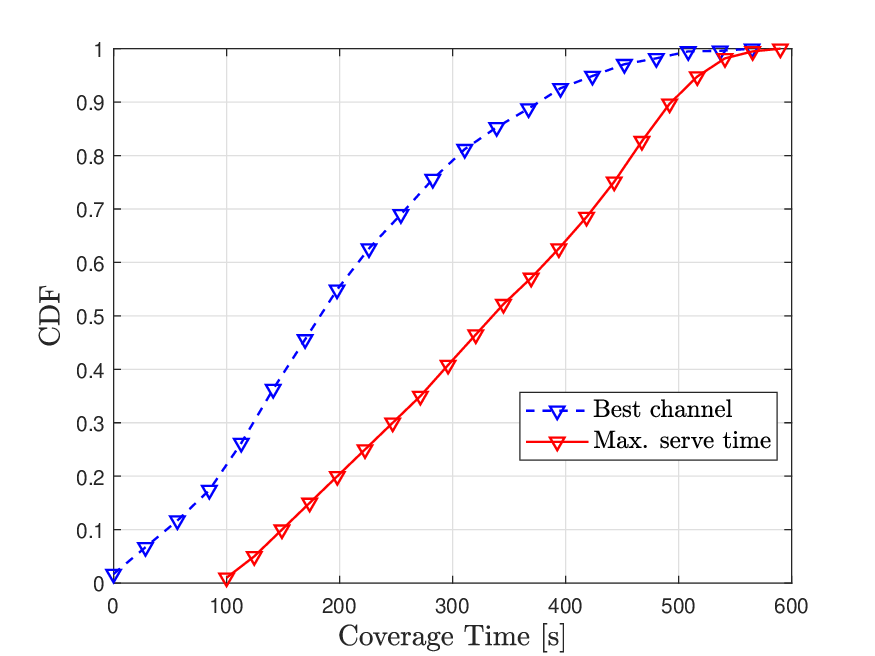}
    \caption{Per-user coverage time  without RSAP handover when the RSAP is selected based on the best channel and based on the maximum service time.  }
    \label{Fig5}
\vspace{-0.5cm}
\end{figure}

Fig. \ref{Fig5},  plots the CDF of the coverage time, i.e.,  the service time, which is the time in which the user receives services without RSAP handover.  As can be observed in the figure, the maximum service time approach outperforms the best channel in terms of coverage time. This result aligns with expectations, given that the satellite with the best channel doesn't guarantee longer service time. From Figs. \ref{Fig4} and \ref{Fig5}, we can conclude that the proposed UC-DMIMO system can enhance the system performance in terms of both downlink spectral efficiency and handover rate. 

\vspace{-0.4cm}
\section{Conclusion}
In this paper, we proposed a scalable distributed massive MIMO system for LEO satellite networks using dynamic user-centric clustering. The proposed UC-DMIMO system underscores its effectiveness in improving the downlink spectral efficiency and handover rate. Notably, compensating for propagation delay phase shifts significantly enhances system efficiency, either by boosting the downlink data rate or reducing antenna requirements. Each user is assigned an RSAP which shares the user information with the other satellites in its SC. Two distinct methods are introduced for RSAP selection; either opting for the LEO satellite with the best channel or choosing the one with the maximum service time.  The comparison between RSAP selection methods further reinforces the system's superiority over a baseline NCT systems. Moreover, with almost the same spectral efficiency performance as the FC-DMIMO system, the UC-DMIMO provides significantly smaller cluster size which reduces the backhaul overhead requirements. Overall, these findings affirm the potential of the proposed UC-DMIMO architecture to elevate system performance in terms of spectral efficiency,  handover rate, and scalable implementation.
\vspace{-0.3cm}
\bibliographystyle{IEEEtran}
\def\bibfont{\footnotesize}

\begin{thebibliography}{99}
\bibitem{Wangcellular}C.-X. Wang et al., “Cellular architecture and key technologies for
5G wireless communication networks,” \textit{IEEE Commun. Mag.}, vol. 52,
no. 2, pp. 122–130, Feb. 2014.

\bibitem{ArapoglouMIMO}
P.-D. Arapoglou, K. Liolis, M. Bertinelli, A. Panagopoulos, P. Cottis,
and R. De Gaudenzi, “MIMO over satellite: A review,” \textit{IEEE Commun.
Surveys Tuts.}, vol. 13, no. 1, pp. 27–51, Jan. 2011.
 
\bibitem{HeoMIMO} J. Heo, et al. “MIMO satellite communication systems: A survey from the PHY layer perspective,” \textit{IEEE Commun.  Surveys Tuts.}, vol. 25, no. 3, pp. 1543-1570, 2023.

\bibitem{AbdelsadekDistributed} M. Y. Abdelsadek, G. Karabulut Kurt, and H. Yanikomeroglu, “Distributed massive MIMO for LEO satellite networks,” \textit{IEEE Open J. Commun. Soc.}, vol. 3, pp. 2162–2177, Nov. 2022.

\bibitem{RichterDownlink} R. Richter, I. Bergel, Y. Noam, and E. Zehavi, “Downlink cooperative MIMO in LEO satellites,” \textit{IEEE Access}, vol. 8, pp. 213866–213881, Nov. 2020. 

\bibitem{SyangD} S. Yang, D. Wang, L. Liu, B. Wang, and C. Sun, "Distributed multiple LEO satellites cooperative downlink power enhancement transmission scheme based on OTFS," in \textit{IEEE 23rd Int. Conf.  Commun. Technol. (ICCT)}, Wuxi, China, 2023, pp. 1208-1213, 

\bibitem{ArtiData} M. Arti, “Data detection in multisatellite communication systems,” \textit{IEEE Trans. Aerosp. Electron. Syst.}, vol. 56, no. 2, pp. 1637–1644, Apr. 2020.


\bibitem{abdelsadek2023broadband}
M. Y. Abdelsadek, G. Karabulut Kurt, H. Yanikomeroglu, P. Hu, G. Lamontagne, and K. Ahmed, "Broadband connectivity for handheld devices via LEO satellites: Is distributed massive MIMO the answer?,"  \textit{IEEE Open J. Commun. Soc.}, vol. 4, pp. 713-726, Mar. 2023.

\bibitem{XuEnhancement} Z. Xu, G. Chen, R. Fernandez, Y. Gao, R. Tafazolli,  “Enhancement of direct LEO satellite-to-smartphone communications by distributed beamforming,” \textit{ IEEE Trans. Veh. Technol.} (Early access), 2024.

\bibitem{XLiB} X. Li and B. Shang, “An analytical model for coordinated multi-satellite joint transmission system, ” \textit{arXiv preprint}, arXiv:2311.05189v2, Apr. 2024. 
\bibitem{study3GPP}
 “Study on new radio (NR) to support non-terrestrial networks (Release
15),” 3GPP, Sophia Antipolis, France, Rep. TR 38.811, V15.4.0,
Sep. 2020.
\bibitem{JZHANGPerformance} J. Zhang, L. Dai, Z. He, S. Jin, and X. Li, “Performance analysis of mixed-ADC massive MIMO systems over Rician fading channels,” \textit{IEEE J. Sel. Areas Commun.}, vol. 35, no. 6, pp. 1327–1338,
Jun. 2017.
 \bibitem{AbdelsadekBroadBand} M. Y. Abdelsadek, H. Yanikomeroglu, and G. Karabulut Kurt, “Future ultradense LEO satellite networks: A cell-free massive MIMO approach,” in \textit{Proc. IEEE Int. Conf Commun. Workshops (ICC Workshops)}, 2021,
pp. 1–6.

\bibitem{CellfreeLarsson} H. Q. Ngo, A. Ashikhmin, H. Yang, E. G. Larsson, and T. L. Marzetta,
“Cell-free massive MIMO versus small cells,” \textit{IEEE Trans.
Wireless Commun}, vol. 16, no. 3, pp. 1834–1850, Jan. 2017.

\bibitem{JImpacts}J. Li, M. Liu, P. Zhu, D. Wang, and X. You, “Impacts of asynchronous
reception on cell-free distributed massive MIMO systems,” \textit{IEEE Trans.
Veh. Technol.}, vol. 70, no. 10, pp. 11 106–11 110, Oct. 2021.
\bibitem{YanAsynchronous}H. Yan and I. Lu, “Asynchronous reception effects on distributed massive MIMO-OFDM system,” \textit{IEEE Trans. Commun.}, vol. 67, no. 7,
pp. 4782–4794, Jul. 2019.

\bibitem{JAsynchronous}J. Zheng, J. Zhang, J. Cheng, V. C. M. Leung, D. W. K. Ng, and B. Ai, "Asynchronous Cell-Free Massive MIMO With Rate-Splitting," \textit{ IEEE J. Sel. Areas Commun.}, vol. 41, no. 5, pp. 1366-1382, May 2023.

\bibitem{Bjornsonscalable}E. Björnson and L. Sanguinetti, “Scalable cell-free massive MIMO
systems,” \textit{IEEE Trans. Commun.}, vol. 68, no. 7, pp. 4247–4261,
Jul. 2020

 \bibitem{Wegraph}Z. Wu, F. Jin, J. Luo, Y. Fu, J. Shan, and G. Hu, “A graph-based satellite handover framework for LEO satellite communication networks,” \textit{IEEE
Commun Lets.}, vol. 20, no. 8, pp. 1547–1550, May 2016.
\end{thebibliography}

\end{document}